\documentclass[12pt]{article}
\usepackage{amsmath} 
\input{epsf} 
\def\ltap{\raisebox{-.6ex}{\rlap{$\,\sim\,$}} \raisebox{.4ex}{$\,<\,$}} 
\def\gtap{\raisebox{-.6ex}{\rlap{$\,\sim\,$}} \raisebox{.4ex}{$\,>\,$}}

\newcommand\as{\alpha_{\mathrm{S}}} 
 
\def\beq{\begin{equation}} 
\def\eeq{\end{equation}} 
\def\beeq{\begin{eqnarray}} 
\def\eeeq{\end{eqnarray}} 
 
\def\to{\rightarrow}

\def\ptmin{p_{T{\rm min}}}
\def\ptmax{p_{T{\rm max}}}
\def\ptveto{p_T^{\rm veto}}

\begin{document} 

\begin{titlepage}
\renewcommand{\thefootnote}{\fnsymbol{footnote}}
\vspace*{2cm}

\begin{center}
{\Large \bf NNLO predictions for the Higgs boson signal}
\\[0.5cm]
{\Large \bf in the $H\to WW\to l\nu l\nu$ and $H\to ZZ\to 4l$ decay channels}
\end{center}
\par \vspace{2mm}
\begin{center}
{\bf Massimiliano Grazzini}\\

\vspace{5mm}

INFN, Sezione di Firenze,\\
%and Dipartimento di Fisica,Universit\`a di Firenze,\\ 
I-50019 Sesto Fiorentino, Florence, Italy\\

\vspace{5mm}

\end{center}

\par \vspace{2mm}
\begin{center} {\large \bf Abstract} \end{center}
\begin{quote}
\pretolerance 10000

We consider Standard Model Higgs boson production by gluon--gluon fusion in hadron collisions. We present a calculation of the next-to-next-to-leading 
order QCD corrections to the
%signal
cross section
in the $H\to WW\to l\nu l\nu$ and $H\to ZZ\to 4l$ decay channels.
The calculation is implemented in the parton level
Monte Carlo program {\tt HNNLO} and
allows us to apply arbitrary cuts on the final state leptons and the
associated jet activity.
We present selected numerical results
for the
%Higgs
signal cross section
at the LHC,
by using all the nominal cuts proposed for the
forthcoming Higgs boson search.

\end{quote}

\vspace*{\fill}
\begin{flushleft}
January 2008

\end{flushleft}
\end{titlepage}

\setcounter{footnote}{1}
\renewcommand{\thefootnote}{\fnsymbol{footnote}}

\section{Introduction}

The search for the Higgs boson \cite{Hrev}
and the study of its properties (mass, couplings,
decay widths) are at the heart of the LHC physics program.
In this paper we consider the production 
of the Standard Model (SM) Higgs boson
by the gluon fusion mechanism.

The gluon fusion process $gg \to H$, through a heavy-quark (mainly, top-quark) 
loop, is the main production mechanism of the 
SM Higgs boson $H$ at hadron colliders. 
When combined with the decay channels 
$H \to \gamma \gamma$, $H \to WW$ and $H \to ZZ$, 
this production mechanism is one of 
the most important for Higgs boson searches and studies over the entire
range, 100~GeV$\ltap M_H \ltap$1~TeV, 
of Higgs boson mass $M_H$
to be investigated at the LHC \cite{atlas,cms}. 

The dynamics of the gluon fusion mechanism in controlled by 
strong interactions. Detailed studies of the effect of QCD radiative 
corrections are thus necessary 
to obtain accurate theoretical predictions.

At leading order (LO) in QCD perturbation theory, the cross section is proportional
to $\as^2$, $\as$ being the QCD coupling. The QCD radiative corrections to the total cross section have been computed at the
next-to-leading order (NLO)
in Refs.~\cite{Dawson:1990zj,Djouadi:1991tka,Spira:1995rr}
and found to enhance the cross section by about $80-100\%$.
In recent years also
the next-to-next-to-leading order (NNLO) corrections 
\cite{Harlander:2000mg,Catani:2001ic,Harlander:2001is,Harlander:2002wh,Anastasiou:2002yz,Ravindran:2003um} have been computed.
The NNLO effect is moderate and, for a light Higgs,
it increases the NLO cross result by about $15-20\%$.
The effects of a jet veto on the total cross section has also
been studied up to NNLO \cite{Catani:2001cr}.
We recall that all the NNLO results have been obtained by using 
the large-$M_t$ approximation, $M_t$ being the mass of the top quark.

The NNLO results mentioned above are certainly important, but they refer
to situations where the experimental cuts
are either ignored (as in the case of the total cross section) or taken into account
only in simplified cases (as in the case of the jet vetoed cross section).
Generally speaking, the impact of higher-order corrections
may be strongly dependent on the details of the applied cuts and also the shape of the
distributions is typically affected by these details.

The first NNLO calculation that fully takes into account experimental cuts
was reported in Ref.~\cite{Anastasiou:2005qj}, in the case of the decay mode $H\to\gamma\gamma$.
In Ref.~\cite{Anastasiou:2007mz} the calculation was
extended to the decay mode $H\to WW\to l\nu l\nu$.
The calculations of Refs.~\cite{Anastasiou:2005qj,Anastasiou:2007mz}
were performed with the method described in Ref.~\cite{Anastasiou:2003gr},
based on sector decomposition \cite{sector}. Besides Higgs boson production,
the above method has been applied to the
NNLO QCD calculations of $e^+e^-\to 2$ jets \cite{Anastasiou:2004qd},
vector boson production in hadron collisions \cite{DYdiff}, and to the NNLO
QED calculation of the electron energy spectrum in muon decay
\cite{Anastasiou:2005pn}.

In Ref.~\cite{Catani:2007vq} we have presented an independent NNLO calculation
of the Higgs production cross section, including the decay $H\to\gamma\gamma$.
The method is completely different from
that used in Refs.~\cite{Anastasiou:2005qj,Anastasiou:2007mz}.
Our calculation is based on the {\em subtraction method}.

The subtraction method \cite{Ellis:1980wv}
is probably the most popular technique to
handle and cancel infrared singularities in QCD computations at high energy,
and has lead to the formulation of general algorithms \cite{Frixione:1995ms,Catani:1996vz}
to perform NLO calculations in
a relatively straightforward manner, once the relevant amplitudes are available.
In recent years, several research groups have been working
to develop general NNLO
extensions of the subtraction method 
\cite{Kosower,Weinzierl,Frixione:2004is,GGG,ST}.
NNLO results, however,
have been obtained only in some specific processes.
The calculation of $e^+e^-\to 2~{\rm jets}$ \cite{Gehrmann-DeRidder:2004tv,Weinzierl:2006ij}
was the first to be addressed, and, more recently, the computation of
$e^+e^-\to 3~{\rm jets}$ \cite{GehrmannDe Ridder:2007bj,GehrmannDeRidder:2007jk,GehrmannDeRidder:2007hr} has been completed.

The version of the subtraction method
proposed in Ref.~\cite{Catani:2007vq} can be applied to a specific class of processes,
namely, the production of
colourless high-mass systems (lepton pairs, vector bosons, Higgs bosons, 
$\dots$) in hadron collisions. As usual for calculations performed
within the subtraction formalism, the computation can be organized
into a parton level event generator.
The latter feature is particularly useful, since the user can apply the required cuts on the
final state and plot the corresponding distributions in the form of bin histograms.

In Ref.~\cite{Catani:2007vq} we have applied
our method to the computation
of the Higgs production cross section, including the decay $H\to\gamma\gamma$.
In the present paper we extend the calculation of Ref.~\cite{Catani:2007vq}
to the other
important decay modes of the Higgs boson,
namely, $H\to WW\to l\nu l\nu$ and $H\to ZZ\to 4$ leptons, and present
predictions for the Higgs boson signal
that take into account all the realistic experimental cuts
on the final state leptons and the associated jet activity.

The paper is organized as follows. In Sect.~\ref{sec:hnnlo} we describe
our NNLO Monte Carlo program. In Sect.~\ref{sec:results} we present the results of our calculation for the decay modes $H\to WW\to l\nu l\nu$ and $H\to ZZ\to 4l$.
In Sect.~\ref{sec:summary} we summarize our results.

\section{The {\tt HNNLO} Monte Carlo program}
\label{sec:hnnlo}

The numerical program {\tt HNNLO} is a fortran code that implements
the version of the subtraction method proposed in Ref.~\cite{Catani:2007vq}.
The program computes the Higgs boson production cross section
at hadron colliders up to NNLO in QCD perturbation theory.

The cross section up to (N)NLO can be written as
\begin{equation}
\label{main}
d{\sigma}^{H}_{(N)NLO}={\cal H}^{H}_{(N)NLO}\otimes d{\sigma}^{H}_{LO}
+\left[ d{\sigma}^{H+{\rm jets}}_{(N)LO}-
d{\sigma}^{CT}_{(N)LO}\right]\;\; .
\end{equation}
The first term ({\em virtual}) is the simplest to compute numerically:
it contains
the LO cross section $d\sigma_{LO}^H$ at $q_T=0$, $q_T$ being the transverse momentum of the Higgs boson, suitably convoluted with a hard
function ${\cal H}$ which includes the regularized one-loop (two-loop) corrections
to the LO process.
The second term ({\em real}) is the most cumbersome to evaluate. Its
first contribution, $d{\sigma}^{H+{\rm jets}}_{(N)LO}$,
is the (N)LO cross section for the production of the Higgs boson
in association with one (or more) jets.
This contribution is evaluated with the version of the subtraction method of Ref.~\cite{Catani:1996vz}, as implemented
in the MCFM \cite{mcfm} package. When $q_T\to 0$,
$d{\sigma}^{H+{\rm jets}}_{(N)LO}$ is divergent, and
is supplemented with the subtraction of a suitable counterterm,
$d{\sigma}^{CT}_{(N)LO}$.
The difference in the
square bracket of Eq.~(\ref{main}) is thus finite
as $q_T\to 0$.

In the present version of the code (version 1.1) we have implemented
three decay modes for the Higgs boson: $H\to\gamma\gamma$ \cite{Catani:2007vq}, $H\to WW\to l\nu l\nu$
\footnote{Results for this decay channel were presented at the Les Houches Workshop ``Physics at TeV Colliders'' in june 2007, and at the Radcor Conference in october 2007.}
and $H\to ZZ\to 4$ leptons. In the latter case the user can choose between $H\to ZZ\to \mu^+\mu^- e^+e^-$ and $H\to ZZ\to e^+e^-e^+e^-$, which includes the appropriate interference contribution.
The program can be downloaded from \cite{hnnloweb}, together with
some accompanying notes.

\section{Results up to NNLO}
\label{sec:results}

\subsection{Preliminaries}

We consider Higgs boson production at the LHC (e.g. $pp$ collisions at $\sqrt{s}=14$ TeV).  We use MRST2004 parton
distributions \cite{Martin:2004ir},
with densities and $\as$ evaluated at each corresponding order
(i.e., we use $(n+1)$-loop $\as$ at N$^n$LO, with $n=0,1,2$).
Unless stated otherwise, renormalization and factorization scales
are set to their default values, $\mu_R=\mu_F=M_H$.
We remind the reader that the calculation is done in the $M_t\to \infty$ limit.
As for the electroweak couplings, we use the scheme where
the input parameters are $G_F$, $M_Z$, $M_W$ and $\alpha(M_Z)$.
In particular we take $G_F=1.16639\times 10^{-5}$ GeV$^{-2}$, $M_Z=91.188$ GeV, $M_W=80.419$ GeV and $\alpha(M_Z)=1/128.89$.
The decay matrix elements are implemented at Born level, i.e.,
radiative corrections are completely neglected\footnote{We note that the full QCD+EW corrections to the decay modes
$H\to WW(ZZ)\to 4$ leptons have been recently computed \cite{Bredenstein:2006rh}.}.
The Higgs boson is treated in the narrow-width approximation,
but in the $W$ and $Z$ decays we take into account finite width effects,
by using $\Gamma_W=2.06$ GeV and $\Gamma_Z=2.49$ GeV.
As far as jets are concerned, we use the $k_T$-algorithm \cite{ktalg} with jet size $D=0.4$.

\subsection{$H\to WW\to l\nu l\nu$}

We consider the production of a Higgs boson with mass $M_H=165$ GeV.
The width is computed with the program HDECAY \cite{Djouadi:1997yw} to be $\Gamma_H=0.255$ GeV.
With this choice of $M_H$ the Higgs boson decays almost entirely
into $WW$ pairs. We consider the decay $W\to l\nu$ by assuming
only one final state lepton combination.
The corresponding inclusive cross sections are given in Table \ref{tab:wwnocuts}.
The NLO and NNLO $K$-factors are 1.84 and 2.21, respectively, and are in good agreement with the inclusive $K$-factors from the calculation
of the total NLO and NNLO cross section \cite{Harlander:2002wh,Anastasiou:2002yz,Ravindran:2003um}.

\begin{table}[htbp]
\begin{center}
\begin{tabular}{|c|c|c|c|}
\hline
$\sigma$ (fb)& LO & NLO & NNLO\\
\hline
\hline
$\mu_F=\mu_R=M_H/2$ & $136.37\pm 0.09$ & $241.59\pm 0.43$ & $268.7\pm 1.8$\\
\hline
$\mu_F=\mu_R=M_H$ & $112.08\pm 0.07$ & $206.46\pm 0.33$ & $247.2\pm 1.3$ \\
\hline
$\mu_F=\mu_R=2M_H$ & $92.88\pm 0.06$ & $178.43\pm 0.25$ & $227.4\pm 0.8$\\
\hline
\end{tabular}
\end{center}
\caption{{\em Cross sections for $pp\to H+X\to WW+X\to l\nu l\nu+X$ at the LHC
when no cuts are applied.}}
\label{tab:wwnocuts}
\end{table}

We first apply a set of {\em preselection} cuts taken from the study of Ref.~\cite{Davatz:2004zg}.
\begin{enumerate}
\item The event should contain two leptons of opposite charge having $p_T$ larger than 20 GeV and rapidity $|y|<2$;
\item The missing $p_T$ of the event should be larger than $20$ GeV;
\item The invariant mass of the charged leptons should be smaller than $80$ GeV;
\item The azimuthal separation of the charged leptons in the
transverse plane ($\Delta\phi$) should be smaller than $135^o$.
\end{enumerate}
The first cut selects dilepton events originating from
the decay of $W$ or $Z$ bosons.
Lepton pairs originating from the inclusive production
of a $Z$ boson are mostly rejected with cuts 2-4.
The corresponding cross sections are given in Table \ref{tab:wwpresel}.
\begin{table}[htbp]
\begin{center}
\begin{tabular}{|c|c|c|c|}
\hline
$\sigma$ (fb)& LO & NLO & NNLO\\
\hline
\hline
$\mu_F=\mu_R=M_H/2$ & $64.03\pm 0.06$& $113.57\pm 0.28$ & $124.75\pm 1.28$\\
\hline
$\mu_F=\mu_R=M_H$ & $53.10\pm 0.05$ & $97.30\pm 0.21$ & $116.24\pm 0.81$ \\
\hline
$\mu_F=\mu_R=2M_H$ & $44.32\pm 0.04$ & $84.69\pm 0.16$ & $106.48\pm 0.61$\\
\hline
\end{tabular}
\end{center}
\caption{{\em Cross sections for $pp\to H+X\to WW+X\to l\nu l\nu+X$ at the LHC
when preselection cuts are applied.}}
\label{tab:wwpresel}
\end{table}
Comparing with Table~\ref{tab:wwnocuts} we find that the efficiency is $47\%$ both at NLO and at NNLO.
The corresponding NLO and NNLO $K$-factors are 1.83 and 2.19.
With respect to the inclusive case, we notice that the preselection cuts
do not alter significantly the convergence of the perturbative expansion.

\begin{figure}[htb]
\begin{center}
\begin{tabular}{c}
\epsfxsize=10truecm
\epsffile{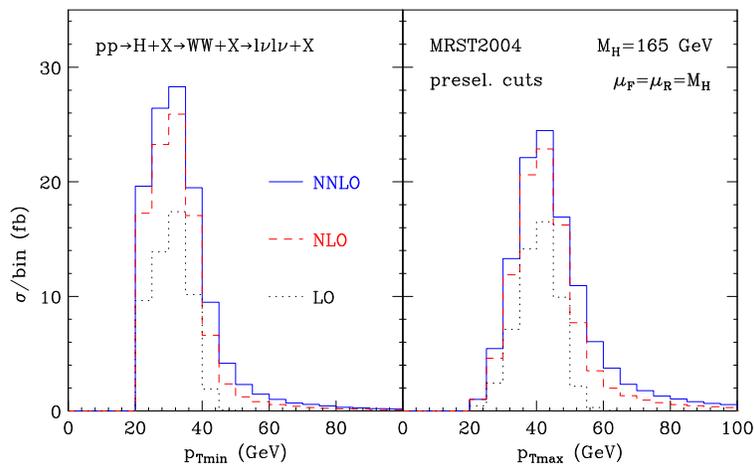}\\
\end{tabular}
\end{center}
\caption{\label{fig:wwpt}
{\em Transverse momentum spectra of the charged leptons for
$pp\to H+X\to WW+X\to l\nu l\nu+X$
at LO (dots), NLO (dashes) and NNLO (solid). Preselection cuts are applied.}}
\end{figure}

\begin{figure}[htb]
\begin{center}
\begin{tabular}{c}
\epsfxsize=10truecm
\epsffile{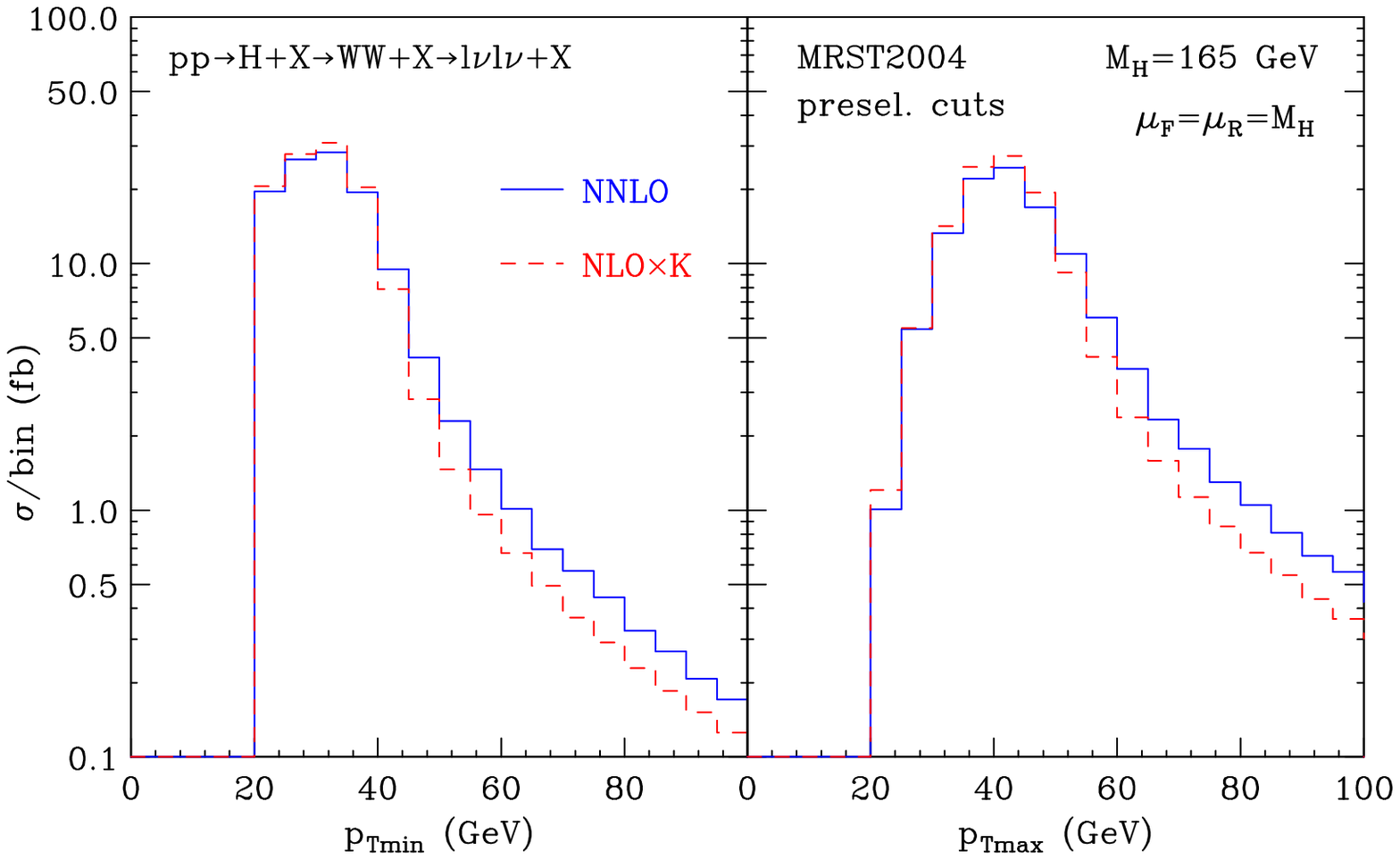}\\
\end{tabular}
\end{center}
\caption{\label{fig:wwptlog}
{\em As in Fig.~\ref{fig:wwpt}:
comparison of $p_T$ spectra at NNLO (solid) with NLO normalized to the same area (dashes).}}
\end{figure}

For each event, we classify the transverse momenta of the charged leptons according to their
minimum and maximum value,  
$\ptmin$ and $\ptmax$.
In Fig.~\ref{fig:wwpt} we plot the $\ptmin$ and $\ptmax$ distribution
at LO, NLO and NNLO.
We see that QCD corrections tend to make the distributions harder.
This can be also appreciated from Fig.~\ref{fig:wwptlog}, where we compare the NNLO distributions with the NLO ones, normalized to the same area.

%%====================================
%%====================================

%%====================================
%%====================================

%%====================================
\begin{figure}[htb]
\begin{center}
\begin{tabular}{c}
\epsfxsize=10truecm
\epsffile{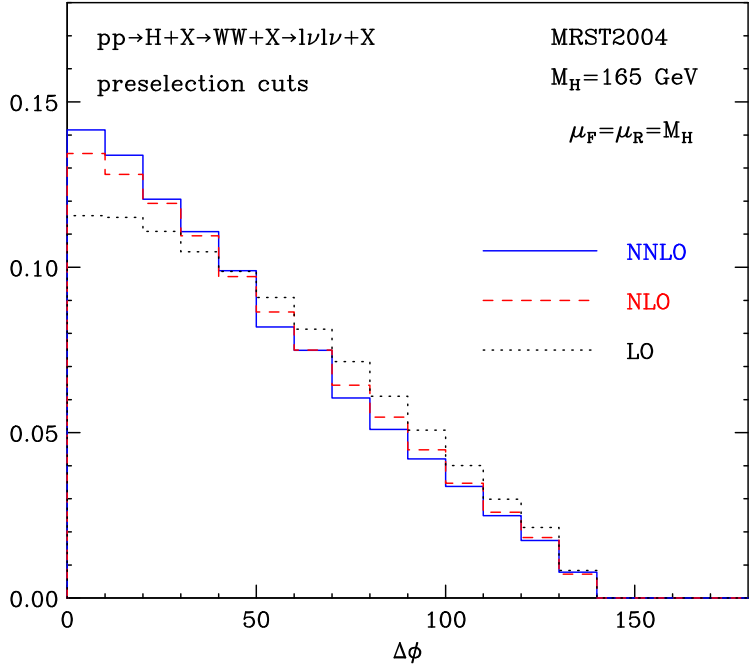}\\
\end{tabular}
\end{center}
\caption{\label{fig:deltaphi}
{\em Normalized distribution in the variable $\Delta\phi$ when preselection cuts are applied.}}
\end{figure}
%%====================================

In Fig.~\ref{fig:deltaphi} we plot the $\Delta\phi$ distribution at LO, NLO and NNLO. As is well known \cite{Dittmar:1996ss},
for the Higgs boson signal the leptons tend to be close in angle,
and thus most of the events are concentrated at small $\Delta \phi$.
We notice that the steepness of the distribution increases
when going from LO to NLO and from NLO to NNLO.
As a consequence, the efficiency of a cut on this variable also
increases
with the perturbative order.

We finally consider the following {\em selection} cuts \cite{Davatz:2004zg},
which are designed to isolate the Higgs boson signal:
\begin{enumerate}
\item The two charged leptons, with rapidity $|y|<2$,
should fulfil $\ptmin > 25$ GeV and\\ $35~{\rm GeV}< \ptmax <50$ GeV;
\item The missing $p_T$ of the event should be larger than 20 GeV;
\item The invariant mass of the charged leptons should be smaller than $35$ GeV;
\item The azimuthal separation of the charged leptons in the
transverse plane ($\Delta\phi$) should be smaller than $45^o$;
\item Finally, there should be no jets with $p_T^{\rm jet}$ larger
than a given value $\ptveto$.
\end{enumerate}
These cuts further exploit:
{\em i)} the shape of the $\ptmin$ and $\ptmax$ distributions shown in Fig.~\ref{fig:wwpt}; {\em ii)} the strong angular correlations
of the charged leptons leading to the steep $\Delta\phi$ distribution
in Fig.~\ref{fig:deltaphi};
{\em iii)} the fact that the decay of top quarks from
the $t{\bar t}$ background produces $b$-jets with large transverse momentum. A jet veto is thus very efficient to suppress this background.
 
In Table \ref{tab:wwsel} we report the corresponding
cross sections in the case of $\ptveto=30$ GeV.

\begin{table}[htbp]
\begin{center}
\begin{tabular}{|c|c|c|c|}
\hline
$\sigma$ (fb)& LO & NLO & NNLO\\
\hline
\hline
$\mu_F=\mu_R=M_H/2$ & $17.36\pm 0.02$ & $18.11\pm 0.08$ & $15.70\pm 0.32$\\
\hline
$\mu_F=\mu_R=M_H$ & $14.39\pm 0.02$ & $17.07\pm 0.06$ & $15.99\pm 0.23$ \\
\hline
$\mu_F=\mu_R=2M_H$ & $12.00 \pm 0.02$ & $15.94\pm 0.05$ & $15.68\pm 0.20$\\
\hline
\end{tabular}
\end{center}
\caption{{\em Cross sections for $pp\to H+X\to WW+X\to l\nu l\nu+X$ at the LHC
when selection cuts are applied and $\ptveto=30$ GeV.}}
\label{tab:wwsel}
\end{table}

A comparison with Table \ref{tab:wwpresel} reveals that
the cross section
is strongly suppressed with respect to the case in which only
preselection cuts are applied: the efficiency turns out
to be $8\%$ at NLO and $6\%$ at NNLO.
The scale dependence of the result is strongly reduced at NNLO,
being of the order of the error from the numerical integration.
The impact of higher order corrections is also
drastically changed. The $K$-factor is now 1.19 at NLO and
1.11 at NNLO.
As expected, the jet veto tends to stabilize the
perturbative expansion.
The latter point has a simple qualitative explanation \cite{Catani:2001cr}. 

It is well known that the effect of higher order contributions to the inclusive Higgs production cross section is large.
The dominant part of this effect is due to soft and virtual
contributions. The characteristic scale of the highest
transverse momentum $p_T^{\rm max}$ of the accompanying jets
is indeed $p_T^{\rm max}\sim\langle 1-z\rangle M_H$,
where $z=M_H^2/{\hat s}$ and $\langle 1-z\rangle$ measures
the average distance from the partonic threshold.
As a consequence, the effect of the jet veto is small unless
$\ptveto$ is substantially smaller than $p_T^{\rm max}$.
Decreasing $\ptveto$, the enhancement of the inclusive cross
section due to soft-radiation at higher orders is reduced,
and the jet veto improves the convergence of
the perturbative series.
Note, however, that when $\ptveto$ is much smaller than the
characteristic scale $p_T^{\rm max}\sim\langle 1-z\rangle M_H$,
the coefficients of the perturbative series contain
logarithmically enhanced contributions that may invalidate
the convergence of the fixed order expansion.

\begin{figure}[htb]
\begin{center}
\begin{tabular}{c}
\epsfxsize=10truecm
\epsffile{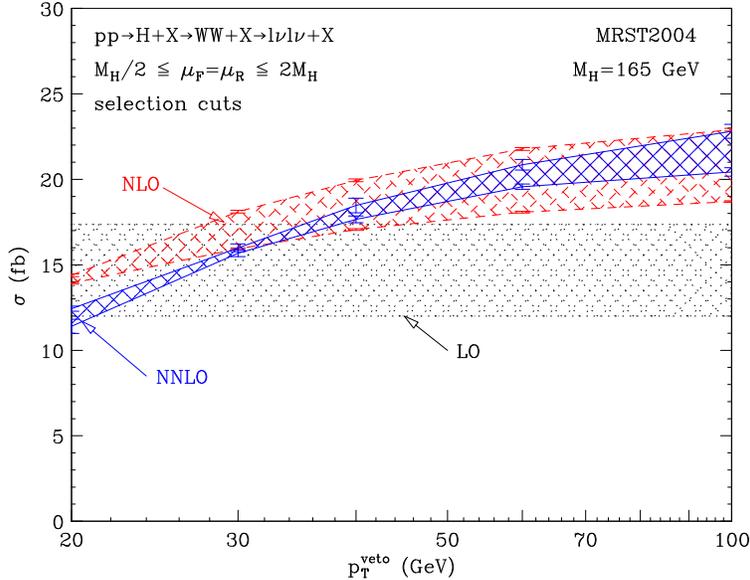}\\
\end{tabular}
\end{center}
\caption{\label{fig:veto}
{\em Cross sections as a function of $\ptveto$ when selection cuts are applied. The bands are obtained
by varying $\mu_R=\mu_F$ between $M_H/2$ and $2M_H$.}}
\end{figure}

In order to estimate the perturbative uncertainties
affecting our calculation,
in Fig.~\ref{fig:veto} we report the LO, NLO and NNLO bands
as a function of $\ptveto$, when all the other selection cuts are applied.
The bands are obtained by
varying $\mu_F=\mu_R$ between $M_H/2$ and $2M_H$.
%%====================================
%%====================================
The results of Fig.~\ref{fig:veto} deserve some discussion.

At LO there are no jets accompanying the Higgs boson,
and thus the cross section is independent on $\ptveto$. The NLO band overlaps with the LO one for $\ptveto$ smaller than about
50 GeV. Without jet veto ($\ptveto\to \infty$)
the $K$-factor, defined with respect to the LO cross section
at central values of the scales, ranges between 1.32 ($\mu_F=\mu_R=2M_H$) and 1.63 ($\mu_F=\mu_R=M_H/2$).
Comparing with the inclusive results, we see that the selection cuts 1-4 alone
already imply a reduction of the impact of higher order corrections.
We also observe that the NLO band becomes very narrow as soon as $\ptveto$ decreases.

The NNLO band overlaps with the NLO one for $\ptveto\gtap 30$ GeV and thus
suggests a good convergence of the perturbative expansion in this region of $\ptveto$.
On the contrary, for $\ptveto\ltap 30$ GeV,
the NNLO band is very narrow
and does not overlap with the NLO one, suggesting that, in this region, the perturbative
uncertainty obtained through scale variations is likely
to be underestimated.

The NNLO corrections to the $pp\to H+X\to WW+X\to l\nu l\nu+X$ at the LHC
were independently computed in Ref.~\cite{Anastasiou:2007mz}.
The preselection cuts we use are the same as those considered in
Ref.~\cite{Anastasiou:2007mz}. Taking into account the different
normalization\footnote{In our calculation we strictly apply the large-$M_t$ approximation, whereas in the calculation of Ref.~\cite{Anastasiou:2007mz}
the results are normalized to the Born cross section with exact top-quark mass dependence.},
the ensuing cross sections in Table \ref{tab:wwpresel}
are in good agreement
with those given in Table 2 of Ref.~\cite{Anastasiou:2007mz}.
When selection cuts are applied, a direct comparison
%with the results of Ref.~\cite{Anastasiou:2007mz}
is not possible,
since the cuts we employ are not exactly the same.
Fig.~1 of Ref.~\cite{Anastasiou:2007mz} shows that, when only the jet veto
is applied, the NLO and NNLO bands computed as in Fig.~\ref{fig:veto}
overlap for $\ptveto\ltap 40$ GeV.
Nonetheless, when all the selection cuts are applied and $\ptveto=25$ GeV,
the NLO and NNLO results reported in Table 3 of Ref.~\cite{Anastasiou:2007mz} do not overlap.
Although the selection cuts we use are not exactly the same,
the latter result is consistent with the behaviour we observe in Fig.~\ref{fig:veto}. In the recent study of Ref.~\cite{Anastasiou:2008ik}
the efficiencies obtained at NNLO are shown
to be in good agreement with those
predicted by the MC@NLO event generator \cite{MCatNLO}.

\subsection{$H\to ZZ\to e^+e^-e^+e^-$}

We now consider the production of
a Higgs boson with mass $M_H=200$ GeV.
The width is computed with the program HDECAY \cite{Djouadi:1997yw} to be $\Gamma_H=1.43$ GeV.
In this mass region the dominant decay mode is $H\to ZZ\to 4l$,
providing a clean four lepton signature.
In the following we
consider the decay of the Higgs boson in two identical lepton pairs.
When no cuts are applied, the signal cross sections
are reported in Table \ref{tab:nocuts}. We find that the interference contribution is smaller than $1\%$ in
this mass region. The ensuing inclusive cross section is thus a factor of 2 smaller than the cross section in the
decay channel $H\to ZZ\to  \mu^+\mu^- e^+ e^-$\footnote{In the case of $H\to ZZ\to e^+e^-e^+e^-$ there is an additional diagram, obtained for example by exchanging the momenta of the two electrons, but there is also a symmetry factor 1/4, due to the
two pairs of identical particles \cite{Zecher:1994kb}.}.

\begin{table}[htbp]
\begin{center}
\begin{tabular}{|c|c|c|c|}
\hline
$\sigma$ (fb)& LO & NLO & NNLO\\
\hline
\hline
$\mu_F=\mu_R=M_H/2$ & $2.457\pm 0.001$ & $4.387\pm 0.006$& $4.90\pm 0.03$\\
\hline
$\mu_F=\mu_R=M_H$ & $2.000 \pm 0.001$ & $3.738\pm 0.004$ & $4.52 \pm 0.02$ \\
\hline
$\mu_F=\mu_R=2M_H$ &$1.642\pm 0.001$ &$3.227\pm 0.003$ & $4.14\pm 0.01$\\
\hline
\end{tabular}
\end{center}
\caption{{\em Cross sections for $pp\to H+X\to ZZ+X\to e^+e^-e^+e^-+X$ at the LHC
when no cuts are applied.}}
\label{tab:nocuts}
\end{table}
The NLO $K$-factor is $K=1.87$ whereas at NNLO we have $K=2.26$.
These results are in good agreement with those obtained
from the calculation
of the total NLO and NNLO cross section \cite{Harlander:2002wh,Anastasiou:2002yz,Ravindran:2003um}.

We consider the following cuts \cite{cms}:
\begin{enumerate}
\item For each event, we order the transverse momenta of the leptons from the largest ($p_{T1}$) to the smallest ($p_{T4}$). They are required to fulfil the following
thresholds:\\ $p_{T1}>30~{\rm GeV}~~~~p_{T2}>25~{\rm GeV}~~~~p_{T3}>15~{\rm GeV}~~~~p_{T4}>7~{\rm GeV}$\,;
\item Leptons should be central: $|y|< 2.5$;
\item Leptons should be isolated: the total transverse energy $E_T$ in a cone of radius 0.2 around each lepton should fulfil $E_T< 0.05~p_T$;
\item For each possible $e^+e^-$ pair, the closest ($m_1$)
and next-to-closest ($m_2$) to $M_Z$ are found.
Then $m_1$ and $m_2$ are required to be $81$ GeV $< m_1 < 101$ GeV and $40$ GeV $< m_2 < 110$ GeV.
\end{enumerate}
These cuts are designed
to maximize the statistical significance for an early discovery,
but to keep the possibility for a more detailed analysis of
the properties of the Higgs boson.
The corresponding cross sections are reported in Table \ref{tab:cuts}.
\begin{table}[htbp]
\begin{center}
\begin{tabular}{|c|c|c|c|}
\hline
$\sigma$ (fb)& LO & NLO & NNLO\\
\hline
\hline
$\mu_F=\mu_R=M_H/2$ & $1.541 \pm 0.002$ & $2.764\pm 0.005$ & $ 3.013\pm 0.023$\\
\hline
$\mu_F=\mu_R=M_H$ & $1.264\pm 0.001$ & $2.360\pm 0.003$ & $2.805\pm 0.015$\\
\hline
$\mu_F=\mu_R=2M_H$ & $1.047\pm 0.001$ & $2.044 \pm 0.003$ & $2.585\pm 0.010$\\
\hline
\end{tabular}
\end{center}
\caption{{\em Cross sections for $pp\to H+X\to ZZ+X\to e^+e^-e^+e^-+X$ at the LHC
when cuts are applied.}}
\label{tab:cuts}
\end{table}

Comparing with Table~\ref{tab:nocuts}, we see that, contrary to what happens in the $H\to WW\to l\nu l\nu$ decay mode,
the cuts are quite mild, the efficiency being $63\%$ at NLO and $62\%$ at NNLO.
The NLO and NNLO $K$-factors are $1.87$ and $2.22$, respectively.
Comparing with the inclusive case, we
conclude that these cuts do not change significantly
the impact of QCD radiative corrections.
We also find that the effect of lepton isolation is mild: at NNLO it reduces the accepted cross section by about $4\%$.

\begin{figure}[htb]
\begin{center}
\begin{tabular}{c}
\epsfxsize=12truecm
\epsffile{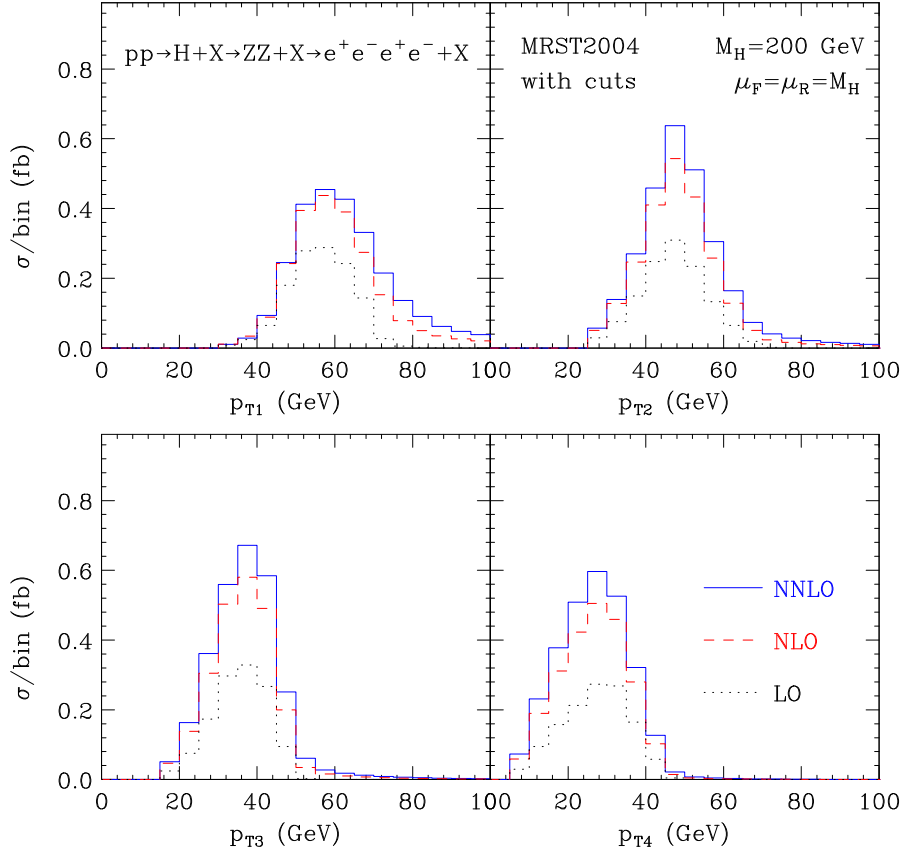}\\
\end{tabular}
\end{center}
\caption{\label{fig:ptlept}
{\em Tranverse momentum spectra of the final state leptons
for $pp\to H+X\to ZZ+X\to e^+e^-e^+e^-+X$,
ordered according to decreasing $p_T$,
at LO (dotted), NLO (dashed), NNLO (solid).}}
\end{figure}

In Fig.~\ref{fig:ptlept} we plot the $p_T$ spectra of the final state leptons.
We note that at LO, without cuts, the $p_{T1}$ and $p_{T2}$ are kinematically bounded by $M_H/2$, whereas $p_{T3}< M_H/3$ and $p_{T4}<M_H/4$.
It is well known that, in the vicinity of kinematical boundaries,
QCD cross sections may develop perturbative instabilities beyond
a given order, if the behaviour of the cross section is not smooth at that order \cite{Catani:1997xc}.
This is what can be observed in the $p_T$ spectra of the photons
in the $H\to\gamma\gamma$ decay mode \cite{Catani:2007vq}.
In the present case, the effect of the cuts further reduces the kinematically allowed region,
but the LO distributions smoothly reach their kinematical boundary,
and we do not observe such perturbative instabilities beyond LO.

\begin{figure}[htb]
\begin{center}
\begin{tabular}{c}
\epsfxsize=12truecm
\epsffile{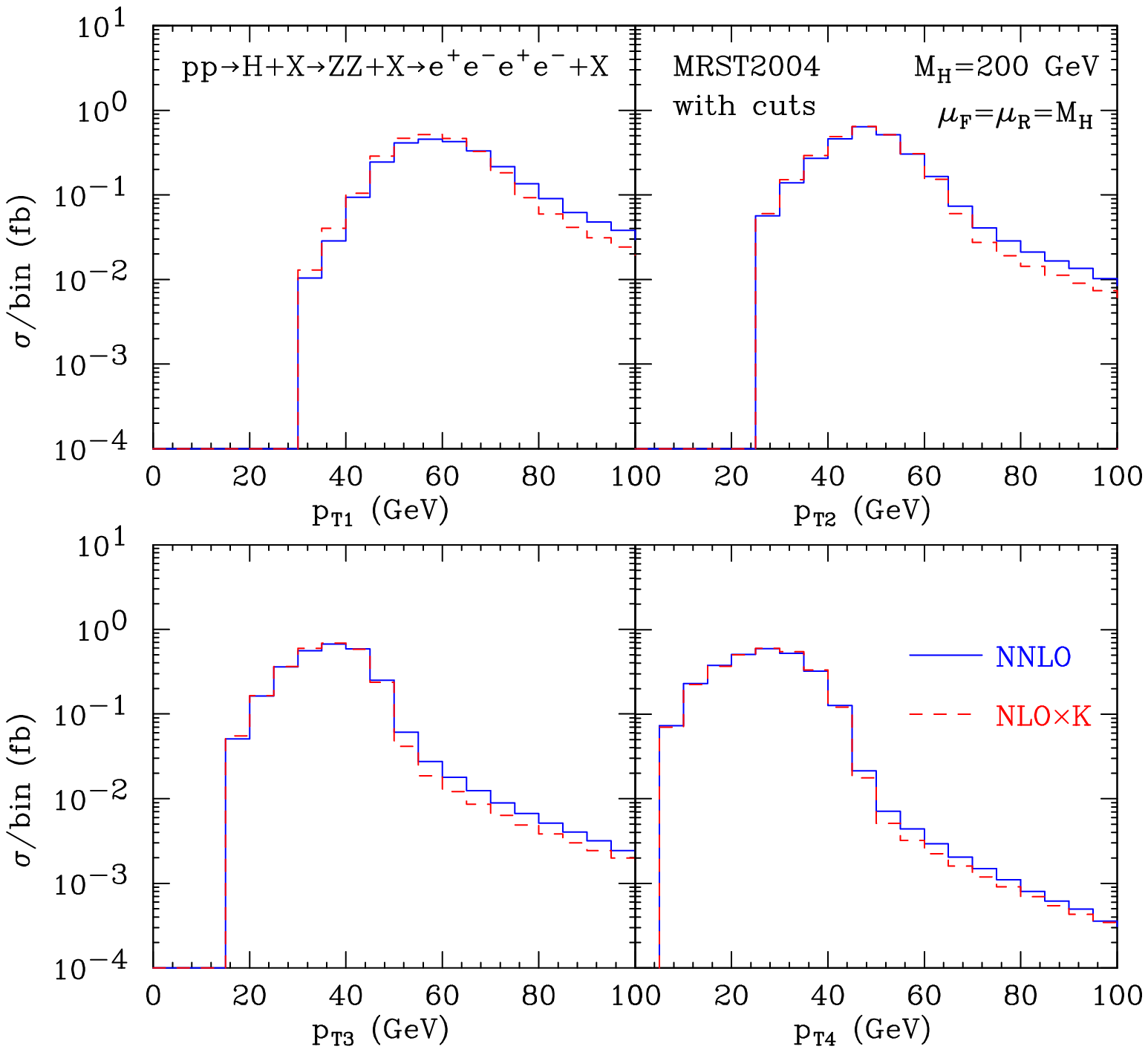}\\
\end{tabular}
\end{center}
\caption{\label{fig:ptlept_log}
{\em As in Fig.~\ref{fig:ptlept}: comparison of lepton $p_T$ spectra at NNLO (solid) with NLO normalized to the same area (dashes).}}
\end{figure}

As in Fig.~\ref{fig:wwpt}, in Fig.~\ref{fig:ptlept} we see that QCD corrections tend to make the distributions harder.
This can be also appreciated from Fig.~\ref{fig:ptlept_log}, where we compare the NNLO distributions with the NLO ones, normalized to the same area.
%%====================================
%%====================================

%%====================================
%%====================================

\section{Summary}
\label{sec:summary}

We have presented a calculation of the NNLO cross section for Higgs
boson production at the LHC, in the decay modes
$H\to WW\to l\nu l\nu$ and $H\to ZZ\to 4$ leptons.
The calculation takes into account all the experimental cuts designed
to isolate the Higgs boson signal \cite{cms,Davatz:2004zg}.
In the case of the decay mode $H\to WW\to l\nu l\nu$, we confirm
previous findings that the effect of radiative corrections is strongly
reduced by the selection cuts.
In the case of the decay mode $H\to ZZ\to 4$ leptons, we find that
the proposed cuts are mild and
do not change dramatically the size of QCD radiative corrections.

Our calculation is implemented in
the numerical program {\tt HNNLO} \cite{hnnloweb}.
The present version of the program includes
the most relevant decay modes of the Higgs boson, namely,
$H\to\gamma\gamma$, $H\to WW\to l\nu l\nu$ and $H\to ZZ\to 4$ leptons.
In the latter case it is possible to choose between
$H\to ZZ\to \mu^+\mu^- e^+e^-$ and $H\to ZZ\to e^+e^-e^+e^-$,
which includes the appropriate interference contribution.
The user can apply all the required cuts on the
final state leptons (photons) and the associated jets
and plot the corresponding distributions in the form of bin histograms.
These features should make our program a useful tool for Higgs studies at the Tevatron and the LHC.

\subsection*{Acknowledgements}
I wish to thank Stefano Catani for helpful discussions and comments.

\end{document}